\documentclass[twocolumn,prl,floatfix,preprintnumbers]{revtex4}
\usepackage{mathrsfs,graphicx,amsmath,amssymb,amsbsy}

\begin{document}

\title{Breached pairing superfluidity: Possible realization in QCD}
\author{Elena Gubankova, W. Vincent Liu and Frank Wilczek}
\affiliation{Center for Theoretical Physics, Department of Physics,
Massachusetts Institute of Technology, Cambridge, MA 02139}

\preprint{MIT-CTP \#3357}
\begin{abstract}


We propose a wide universality class of gapless superfluids, and
analyze a limit that might be realized in quark matter at intermediate
densities.  In the breached pairing color superconducting phase heavy
$s$-quarks, with a small Fermi surface, pair with light $u$ or $d$
quarks.  The groundstate has a superfluid and a normal Fermi component
simultaneously.  We expect a second order phase transition, as a
function of increasing density, from the breached pairing phase to the
conventional color-flavor locked (CFL) phase.

\end{abstract}
\maketitle


Because the primary one-gluon exchange interaction between
high-momentum quarks is attractive for quarks in the color
antisymmetric $\bar{\bf 3}$ channel, it is a firm prediction of QCD that
cold dense quark matter is a color superconductor. At asymptotic
densities (ignoring the $c,b$ and $t$ quarks) the ground state is well
understood: quarks of all three flavors $u,d,$ and $s$, pair according
to the BCS mechanism, forming the color-flavor locked (CFL)
phase~\cite{Alford:99}. 

It is much less clear 
what QCD predicts for the ground state at subasymptotic densities,
which could be relevant for describing neutron stars.  Differences
among the quark masses cause mismatches among the Fermi surfaces of the
species which potentially pair.  There is no longer an abundance
of degenerate low-energy particle-particle or hole-hole states with opposite
momenta both near their Fermi surfaces, and so it is less obvious
what modes are the best candidates for coherent alignment
by attractive interactions.

One much-discussed possibility is the  LOFF phase
(Larkin-Ovchinnikov-Fulde-Ferrel \cite{LOFF:65+64}).  
In the context of QCD these
ideas lead to the crystalline color superconductivity \cite{CSC:02}.  
Here we suggest quite a different
possibility for ordering with mismatched Fermi surfaces, that might be
realized at intermediate densities in QCD.

In QCD at intermediate densities, $\mu\sim 200-300$ MeV, there
is not only mismatch in quark Fermi surfaces but also a different
dispersion relation, since the heavy $s$-quark, unlike the
light $u$ and $d$ quarks, needs not be ultra-relativistic.  This
difference makes plausible a pairing phase, wherein strange quarks are
raised to higher kinetic energies to exploit favorable possibilities
for correlation energy through pairing.  The
opposite limit --- pairing of a heavy and light species when heavy
species has the larger Fermi surface --- is the arena for the interior
gap phase discussed in \cite{Liu-Wilczek:03}.


For illustrative purposes we analyze a toy model 
with a massive $s$ and a massless  
$u$ quark.  The Fermi momenta are related to
chemical potentials as 
\begin{equation}
p_{F}^{u} =\mu-\delta\mu^e\,, \qquad
p_{F}^{s} =\sqrt{(\mu+\delta\mu^e)^{2}-m_{s}^2}
\end{equation}
where $\delta\mu^e$ will be tuned to enforce number equality (a
stand-in for electric neutrality). 
We are interested in the case when the Fermi momentum for the
$s$-quark is smaller than that for the $u$-quark,
$p_{F}^{s}<p_{F}^{u}$. For
simplicity, we shall linearize the $s$ quark
dispersion near its Fermi surface.  
We have checked that this simplification does not alter our results
qualitatively.
Our simplified model then has 
dispersion relations
\begin{equation}
\varepsilon_{{\bf p}}^{u} =V^{u}(p-p_{F}^{u}) \,, \qquad 
\varepsilon_{{\bf p}}^{s} =V^{s}(p-p_{F}^{s}) \label{eq:1.1}
\end{equation}
with  $V^{s} < V^{u}$.

In promoting particles of the heavy species to pair around the large
Fermi surface of light species, there are two competing energetic
factors to consider. 
These are the single-paticle energy cost of such promotion,
$V^{s}|p_{F}^{u}-p_{F}^{s}|$ per pair, 
versus the gain from creating a pair, $\kappa(V^{u}+V^{s})$, where
$\kappa$ is the momentum gap.  
There is a net profit when
\begin{eqnarray}
|p_{F}^{u}-p_{F}^{s}|<\kappa \frac{V^{u}+V^{s}}{V^{s}}
\,.\label{eq:1.2}\end{eqnarray}
For flat dispersion for the $s$-quark, $V^{s}\ll V^{u}$,
promotion of $s$-quarks to a higher $u$-quark Fermi 
surface does not cost much energy,
and a paired phase is favored.
Translating this into possible effects of non-zero $s$-quark mass
the allowed range for pairing is given by
\begin{equation} 
|\delta\mu^e-\frac{m_{s}^{2}}{4\mu }|<\Delta \frac{V^{u}}{V^{s}}\,.
\end{equation}

It can be more economical to promote heavy particles to higher Fermi
momentum than to equalize the two Fermi surfaces by deforming both.   
In earlier
work~\cite{Liu-Wilczek:03} we suggested the terminology ``interior
gap'' for this phenomenon, motivated by the peaking of the gap
parameter interior to the large Fermi surface that carries most of the
spectral weight for gapless modes.  Following that thought, it would be
natural to call the limit we are analyzing here ``exterior gap''.
But to emphasize the common physical
mechanism, we use ``breached pairing'' to cover both.  
The pairing is breached, in that it vanishes in
a solid annulus in momentum space.  

To allow quantitative estimates in analytic form we consider a
schematic model involving pairing of $u$ and $s$ quarks.
Starting with non-interacting degenerate Fermi gases of $u$ and $s$-quarks,
turn on a weak attractive interaction between light and heavy species 
with a coupling $-g<0$.
In a basis of light particles and heavy holes 
the quadratic part of the Hamiltonian is
\begin{equation}
{\cal H}_{\rm quad} = 
\left(\begin{array}{cc}
\psi^{\dagger}_{u{\bf p}}&\psi_{s-{\bf p}}
\end{array}\right)
\left(\begin{array}{cc}
\varepsilon_{{\bf p}}^{u}&-\Delta^{\ast}\\
-\Delta&-\varepsilon_{{\bf p}}^{s}\end{array}\right)
\left(\begin{array}{c}
\psi_{u{\bf p}}\\\psi^{\dagger}_{s-{\bf p}}
\end{array}\right)
\,,\label{eq:1.3}
\end{equation}
where 
the gap parameter is defined as $\Delta = g(2\pi)^{-3} \int
d^3{\bf p}
\langle \psi^{\dagger}_{u{\bf p}}\psi^{\dagger}_{s-{\bf p}}
\rangle_{BP}$, where the groundstate  is assumed to be the breached pairing
state.
${\cal H}_{\rm quad}$ can be diagonalized by 
the Bogoliubov transformation 
\begin{equation}
\left(\begin{array}{c}
\psi_{u{\bf p}}\\\psi^{\dagger}_{s-{\bf p}}
\end{array}\right)=
\left(\begin{array}{cc}
\cos\theta_{{\bf p}}&-\sin\theta_{{\bf p}}\\
\sin\theta_{{\bf p}}&\cos\theta_{{\bf p}}\end{array}\right)
\left(\begin{array}{c}
\tilde{\psi}_{u{\bf p}}\\ \tilde{\psi}^{\dagger}_{s-{\bf p}}
\end{array}\right)
\end{equation}
with $\sin2\theta_{{\bf p}}=\Delta/
\sqrt{\varepsilon^{+\,2}_{{\bf p}}+\Delta^{2}}$. The new fermion
fields $\tilde{\psi}$ define the form of the interior gap
wavefunction  up to a relative phase (due to degeneracy) by  
$\tilde{\psi}|0\rangle_{BP}=0$.  
In the meantime, we obtain two branches of
quasi-particle excitations with the spectra
$E^{\pm}_{{\bf p}}=\varepsilon^{-}_{{\bf p}}\pm
\sqrt{\varepsilon^{+\,2}_{{\bf p}}+\Delta^{2}}$,
where $\varepsilon_{{\bf p}}^{\pm}=\frac{1}{2}
(\varepsilon_{{\bf p}}^{u}\pm \varepsilon_{{\bf p}}^{s})$. 
The order parameter is related to $\theta_\mathbf{p}$ by
$\langle\psi^{\dagger}_{u{\bf p}}\psi^{\dagger}_{s-{\bf p}}\rangle_{BP}
=\frac{1}{2}\sin2\theta_{{\bf p}}$ outside the breached region.

The energy of the breached pairing state $|0\rangle_{BP}$ can be 
computed from the diagonalized Hamiltonian as
\begin{equation}
\langle H \rangle_{BP}=\frac{1}{g}\Delta^{2}-\int_{D}
\frac{d^3{\bf p}}{(2\pi)^{3}}
\left(\varepsilon_{{\bf p}}^{-}+\sqrt{
\varepsilon^{+\,2}_{{\bf p}}+\Delta^{2}}\right)\,.
\end{equation}
The first term is the mean field potential and the integration 
is restricted to the area 
$D$ defined by $(p_{0}-\lambda)\leq |{\bf p}|\leq p_{\Delta}^{-}$ and
$p_{\Delta}^{+}\leq|{\bf p}|\leq (p_{0}+\lambda)$, where 
$p_{\Delta}^{\pm}$ are the two roots of $E^{+}_{{\bf p}}=0$, 
$p_{0}$ is the momentum defined by $\varepsilon_{{\bf
    p}_{0}}^{+}=0$: 
\begin{equation}
p_{0} 
=p_{F}^{u}-\delta p_{F}\frac{V^{s}}{V^{u}+V^{s}} \ ,
\end{equation} and $\lambda$ is the ultraviolet cutoff.
(We will always be assuming $p_0 \gg \lambda \gg \Delta, \delta p_F$.)
Thus  
\begin{equation} \label{e:pForms}
p_{\Delta}^{\pm} 
= p_{F}^{u}-\frac{1}{2}\delta p_{F}\pm\frac{1}{2}
\sqrt{\delta p_{F}^{2}-\frac{4\Delta^{2}}{V^{u}V^{s}}}
\end{equation}
with $\delta p_{F} \equiv p_{F}^{u}-p_{F}^{s}>0$.
Varying the groundstate energy
with respect to $\Delta$, $d\langle H \rangle_{BP}/d\Delta =0$, we
find the integral equation for the gap parameter
\begin{eqnarray}
1 =g \int_{D}\frac{d^3 {\bf p}}{(2\pi)^{3}}\,
\frac{1}{2}\frac{1}{\sqrt{\varepsilon^{+\,2}_{{\bf p}}+\Delta^{2}}}
\,.\label{eq:1.4}  
\end{eqnarray}
Note that the integrand peaks
near to $|\mathbf{p}| = p_0$ where  the energy denominator
vanishes for $\Delta\rightarrow 0$.

We can illuminate the physical meaning of $p_{\Delta}^{\pm}$.  Pairing
of $u$ and $s$ quarks produces two quasiparticle excitation
branches. While $E^-_\mathbf{p}<0$  always holds, the $E^{+}_\mathbf{p}$ branch
changes  sign in the region $[p_{\Delta}^{-},p_{\Delta}^{+}]$.
The negativity of $E^+_\mathbf{p}$ means that the corresponding 
states are singly occupied by, in our chosen particle-hole basis, the 
$u$-quarks.  Since this region has single occupation, it does not
contribute to the pairing interaction, and it must therefore be
excluded from the gap equation integral.  The same phenomenon appears
in a more formal treatment with Green functions \cite{Schaefer:99}.
Integration over energy yields a non-zero contribution only if the
two poles of the Bogoliubov quasiparticles reside in different half
planes of the whole complex energy plane.

Two cases can arise, depending on whether the $E^{+}_\mathbf{p}$ branch can
intersects the zero energy axis:


\paragraph{\it BCS superfluidity.}
For $\delta p_{F}\leq \frac{2\Delta}{\sqrt{V^{u}V^{s}}}$,
$E_{{\bf p}}^{+}$ never intersects $E_{{\bf p}}^{+}=0$.  
Then $E_{{\bf p}}^{+}>0$ and $E_{{\bf p}}^{-}<0$ for all
$\mathbf{p}$.  Pairing is possible in the whole range
$(p_{0}-\lambda)\leq |{\bf p}|\leq (p_{0}+\lambda)$.
We arrive at the BCS solution for the gap
\begin{equation}
\Delta=\lambda(V^{u}+V^{s})\,\exp\left({-\frac{1}{N_{+}(0)g}}\right)
\,,\label{eq:1.5}
\end{equation}
where $N_{+}(0)= {1\over (2\pi)^3}\int d^3 \mathbf{p}\, 
\delta(\epsilon^+_\mathbf{p})$ 
is the density
of states, 
and $\lambda$ is the ultraviolet cutoff. Since $N_{+}(0)\sim
p_{0}^{2}$, the gap parameter  
decreases as a function of increasing $\delta p_F$ with
$p_{F}^{u}$ fixed.

\paragraph{\it Breached pairing (BP) superfluidity.}
This occurs when
$\delta p_{F} > \frac{2\Delta}{\sqrt{V^{u}V^{s}}}$. 
$E_{{\bf p}}^{+}=0$ has two roots.   
We integrate below $p_{\Delta}^{-}$ and above $p_{\Delta}^{+}$, but
exclude the region   
$p_{\Delta}^{-}\leq |{\bf p}|< p_{\Delta}^{+}$. 
A solution of the gap equation exists
only if $g>g_{c}$, with the critical coupling constant
\begin{equation}
g_{c} =\left[{N_{+}(0)}
{\ln\left(\frac{\lambda}{\delta p_{F}} 
\frac{(V^{u}+V^{s})}{\sqrt{V^{u}V^{s}}}\right)}\right]^{-1}
\,.\label{eq:1.6}
\end{equation}
For ${V^{s}}/{V^{u}}\ll 1$ with fixed Fermi momenta,
$g_{c}\rightarrow 0$ becomes arbitrarily weak.  


Expanding the gap equation (\ref{eq:1.4})
for small $\Delta  < \delta p_F$,
we find the critical behavior 
\begin{equation}
\Delta \sim (g-g_{c})^{\frac{1}{2}}
\end{equation}
with $g_{c}$ defined in Eq.~(\ref{eq:1.6}).

The quasiparticle excitations of the breached pairing phase differ
qualitatively from those of BCS. In the BCS phase, 
$E^\pm_\mathbf{p}$ branches are both gapped with their minimum
energies given by 
$|\epsilon^-_{p_0}
\pm 2\Delta\sqrt{V_u V_s}/(V_u+V_s)|\neq 0$, respectively.  
By contrast, in the breached pairing phase $E_\mathbf{p}^{+}$ 
intersects zero energy. Only the
$E_{p}^{-}$ branch is fully
gapped.

\begin{figure}[htbp]
\includegraphics[width=\linewidth]{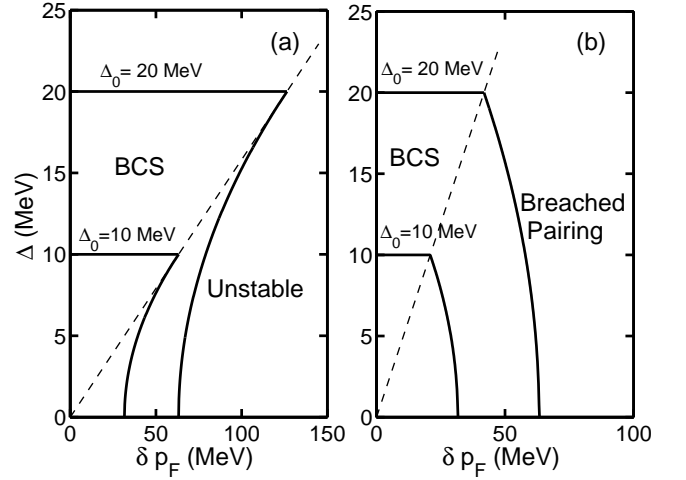}
\caption{Solution of the gap equation as a function of the Fermi
momentum mismatch 
$\delta p_{F}$ for different coupling constants parameterized by 
$\Delta_{0}$. 
(a) Chemical potentials fixed: 
The dashed line, $\delta p_{F}=\frac{2
\Delta}{\sqrt{V^{u}V^{s}}}$, marks the transition between  the BCS
and breached pairing phases.  The range for the unstable breached phase is 
$\frac{\Delta_{0}}{\sqrt{V^{u}V^{s}}}\leq\delta p_{F}\leq
\frac{2\Delta_{0}}{\sqrt{V^{u}V^{s}}}$.  
(b) The total density fixed: 
The dashed line, ploted by $\delta p_{F}=\frac{2\Delta}{\sqrt{V^{u}V^{s}}}
(1-\alpha^2)$,
marks a continuous transition between  the BCS
and breached pairing phases. The range for breached pairing is
defined by $\frac{2\Delta_{0}(1-\alpha^{2})}
{\sqrt{V^{u}V^{s}}}\leq\delta p_{F}\leq
\frac{\Delta_{0}}{\sqrt{V^{u}V^{s}}}$ .  Both (a) and (b) are plotted for
$V^{u}=1$ and $V^{s}=0.1$. }
\label{fig:gapa} 
\end{figure}

For finite $\Delta$, the gap equation 
(\ref{eq:1.4}) needs to be solved self-consistently,
implementing 
the restriction in integration region, $|{\bf p}|\leq p_{\Delta}^{-}$ and 
$|{\bf p}|\geq p_{\Delta}^{+}$, which depends on $\Delta$.
To this end, we transform the integral gap equation into
an algebraic one by introducing a (BCS) gap paramter, $\Delta_{0} \equiv
\Delta(\delta p_{F}=0)$. The BCS gap $\Delta_{0}$
implicitly defines the interaction strength, 
$1/g=N_{+}(0)\ln\left[\lambda(V^{u}+V^{s})/\Delta_{0}\right]$. We
shall use $\Delta_0$ to parameterize 
the interaction strength below. 
Performing the integrals
\begin{eqnarray}
&& 
\int_{p_{0}-\lambda}^{p_{0}+\lambda}
\frac{ dp}{\sqrt{\varepsilon^{+\,2}_{{\bf p}} +\Delta_{0}^{2}}}\nonumber \\
 &=&
\int_{p_{0}-\lambda}^{p_{0}+\lambda} 
\frac{dp}{\sqrt{\varepsilon^{+\,2}_{{\bf p}}+\Delta^{2}}}
-\int_{p_{\Delta}^{-}}^{p_{\Delta}^{+}}
\frac{dp}{\sqrt{\varepsilon^{+\,2}_{{\bf p}}+\Delta^{2}}}
\end{eqnarray}
where 
$\varepsilon^{+}_{{\bf p}}=\frac{1}{2}(V^{u}+V^{s})(p-p_{0})$, 
we obtain
\begin{equation} 
\frac{\Delta_{0}^{2}}{\Delta^{2}}= 
\frac{\varepsilon^{s}_{p_{\Delta}^{+}}}
     {\varepsilon^{s}_{p_{\Delta}^{-}}} 
\end{equation}
for $\lambda \rightarrow \infty$.
After some elementary algebra,  the above equation reduces to 
\begin{equation}
\frac{\Delta_{0}^{2}}{\Delta^{2}}=\frac{\delta p^{}_{F}+\sqrt{\delta p_{F}^{2}
-\frac{4\Delta^{2}}{V^{u}V^{s}}}}{\delta p^{}_{F}-\sqrt{\delta p_{F}^{2}
-\frac{4\Delta^{2}}{V^{u}V^{s}}}}  \,.\label{eq:gap}
\end{equation}
It has a (stable) solution only if  
\begin{equation}
\delta p_F<\frac{\Delta_0}{\sqrt{V^uV^s}}
\,,\label{eq:range}
\end{equation}
which is equivalent to $g>g_c$ (see Eq.~(\ref{eq:1.6})).
In solving Eq.~(\ref{eq:gap}), we distinguish three cases: 

\paragraph{\it I: Chemical potentials fixed.}  Then $\delta p_F$ remains
unchanged before and after pairing, and  
the gap equation permits two solutions:
\begin{equation}
\textstyle
\Delta=\left\{\begin{array}{ll}
\Delta_0\,,  &\mbox{(BCS)} \\
\left[ \Delta_{0} \sqrt{V^{u}V^{s}} \
\big(\delta p_{F}-\delta p_\mathrm{th}\big)\right]^{\frac{1}{2}}
\,,  & \mbox{(unstable BP)}
\end{array}\right.
\label{eq:gap1}
\end{equation}
where $\delta p_\mathrm{th}=
\frac{\Delta_{0}}{\sqrt{V^{u}V^{s}}}$ (threshold value). 
In this case, the BP state 
always has higher energy than the normal state, so it is unstable (see
Eq.~(\ref{eq:CE})).
Fig.~\ref{fig:gapa}a
makes it further 
clear that even if it is a metastable solution at $\delta p_{F} >
\delta p_\mathrm{th}$, 
the BP state always has a smaller $\Delta$ than BCS.  It
is therefore energetically disfavored.   This case was considered by
Sarma and others~\cite{Sarma:63+TI:69} a long time ago.

\paragraph{\it II: Overall quark density fixed.} In this case one
has to adjust the overall chemical potential, 
$\mu_{oa}\equiv {1\over 2} (V^{u}p_{F}^{u}+V^{s}p_{F}^{s}) ={1\over 2}
(V^u+ V^s) p_0$, in order to
accommodate a fixed overall density when pairing occurs.  
We hold the relative
chemical potential, $\delta \mu = V^{u}p_{F}^{u} - V^{s}p_{F}^{s}$ fixed.  The
mismatched Fermi momentum is then altered, $\delta p_F \rightarrow
\delta\tilde{p}_F$, with `$\sim$' denoting 
the new Fermi momenta $\tilde{p}_F^{u,s}$ after pairing. 

We find $\delta\tilde{p}_F$  is determined by
\begin{equation}
p_{0}^{3}-p_{0}^{2}\,\frac{\varepsilon^{\,-}_{p_{F}^{u}}
-\varepsilon^{\,-}_{p_{F}^{s}}}{V^{u}+V^{s}}=
\tilde{p}_{0}^{3}-\tilde{p}_{0}^{2}\,
\frac{\tilde{\varepsilon}^{\,-}_{\tilde{p}_{\Delta}^{+}}
-\tilde{\varepsilon}^{\,-}_{\tilde{p}_{\Delta}^{-}}}{V^{u}+V^{s}}
\,,\label{eq:dpF1}
\end{equation}
where 
$\tilde{p}_{0}=p_{0}+\frac{2V^{u}V^{s}}{V^{u\,2}-V^{s\,2}}
(\delta p_{F}-\delta\tilde p_{F})\,.
$ 
For large Fermi momenta compared to the mismatch in Fermi surfaces,
$\delta p_{F}\ll p_{F}^{u}, p_{F}^{s}$, we find
\begin{equation}
\delta\tilde{p}_{F}=\frac{\delta p_{F}}{1-2\alpha^{2}
\frac{\Delta^{2}}{\Delta_{0}^{2}+\Delta^{2}}}
\,,\label{eq:dpF2}
\end{equation}
where $\alpha \equiv (V^{u}-V^{s})/(V^{u}+V^{s})$.
The gap equation (\ref{eq:gap}) (with $\delta p_F
\rightarrow \delta \tilde{p}_F$) and Eq.~(\ref{eq:dpF2}) 
are solved by
\begin{equation}
\Delta=\left\{\begin{array}{ll}
\Delta_0\,, & \mbox{(BCS)} \\
\left[\Delta_0 \sqrt{V^{u}V^{s}}
\Big({\delta p_\mathrm{th}-\delta p_{F}\over 
2\alpha^{2}-1}\Big)\right]^{1\over 2}\,. &
\mbox{(BP)} \\
\end{array}\right. \label{eq:gap2}
\end{equation}
The breached pairing phase is realized for
$2\alpha^{2}-1>0$, i.e., 
$V^{s}/V^{u}<3-2\sqrt{2}$.  (See Fig.~\ref{fig:gapa}b.) 
The dashed line shows a continuous phase
transition between BCS ($\delta p_F \leq \delta p_F^c$) and breached pairing 
($\delta p_F > \delta p_F^c$), 
where  $\delta p_F^c = \frac{2\Delta_0}{\sqrt{V^{u}V^{s}}}
(1-\alpha^2)$.  There is no overlap region where
both states are solutions.  The breached pairing state
is stable and has lower energy than that of the unpaired matter (see
(\ref{eq:CE}) below).

For higher values of
$V^s/V^u$ (but smaller than one),   the BP solution becomes unstable, being 
of the same class as
(\ref{eq:gap1}). Very close to
$2\alpha^2-1=0$ one finds from
Eq.~(\ref{eq:gap2}) that $\delta p_F^c$ approached $\delta
p_\mathrm{th}$.  The value
$3-2\sqrt{2}$ is model dependent, and will be
different when a more realistic model and a more accurate
approximation scheme are used.
In the limit ${V^{s}}/{V^{u}}\rightarrow 0$ (so $\alpha\rightarrow
1$), the gap equation (\ref{eq:gap})
exhibits the breached pairing phase for all $\delta p_F>0$; BCS
occurs only at the point $\delta p_F=0$.

In mean field theory there is a single second-order phase transition
between the BCS and BP phases (Fig.\ref{fig:gapa}b).  At
fixed mismatch in Fermi momenta $\delta p_{F}$ and as $\Delta_{0}$
increases, one first encounters BP and then the BCS
phase. The transition between the BP and BCS phases occurs
at $\delta p_{F}=\delta p_F^c$.
Since (\ref{eq:gap2}) is symmetric with
respect to interchange of $V^{u}$ and $V^{s}$, the expression for
critical points holds
equally for both  interior and ``exterior'' gap limits.

\paragraph{III: Relative quark density fixed.} In this case, we also
hold the overall chemical potential fixed while allowing the overall
density to vary.  
In the limit $p_0 \gg \lambda\gg \Delta,\delta p_F$, we find 
\begin{equation}
\delta \tilde{p}_F = \sqrt{\delta p_F^2 +{4 \Delta^2\over V^u V^s}}
\,.
\end{equation}
The gap equation (\ref{eq:gap}) (with $\delta p_F
\rightarrow \delta \tilde{p}_F$) has no BCS but BP solution for any
$\delta p_F>0$,
\begin{equation}
\Delta=
\left[ \Delta_{0} \sqrt{V^{u}V^{s}} \
\big(\delta p_\mathrm{th}-\delta p_{F}\big)\right]^{\frac{1}{2}}
\,.\qquad  \mbox{(BP)}
\label{eq:gap3}
\end{equation}
The BCS solution only occurs at a single point, $\delta p_F=0$. 

\paragraph{Condensation energy.}
To leading order in ${\Delta}/\delta p_F$,
the energy difference between the breached pairing
and normal ($\Delta=0$) phases is
\begin{equation}
E_{BP}-E_{N}
= \left\{ \begin{array}{ll}
+{N_+(0) \Delta^4\over 2 \delta p_F^2} {1\over V^u V^s} \,, & \mbox{(I)} \\
- {N_+(0) \Delta^4\over 2 \delta p_F^2} \left( {2\alpha^2 -1
\over {V^u} {V^s}}\right), & \mbox{(II)}  \\
-{N_+(0) \Delta^4\over 2 \delta p_F^2} {1\over V^u V^s}\,. & \mbox{(III)}
\end{array} \right.  \label{eq:CE}
\end{equation}
The BP state is energetically favorable
in case II if $V^s/V^u < 3-2\sqrt{2}$  
(with overall density and $\delta\mu$ 
fixed), and it is always favorable in case III (with 
the relative density and overall $\mu$ fixed).  


In the breached pairing scenario pairing occurs at zero total mometum and
peaks near the light species Fermi surface. It resembles standard BCS
in that the pairing occurs at all angles,  there is no necessity
for translation or rotation symmetry breaking, and there is a 
branch of gapped excitations.  There are,
nevertheless, also gapless modes at two new ``effective'' Fermi
surfaces, between which pairing is suppressed.  
For small $\Delta$ the
energy gain of the breached pairing phase with respect to unpaired normal
matter is $\sim \Delta^{4}$, parametrically less than $\sim \Delta^{2}$ in the
BCS case.

\paragraph{Discussion.}
In the context of QCD, enforcing charge neutrality will make case III
relevant for the $u$ and $d$ quarks.  For pairing between strange and
light quarks, case II is relevant.  Thus, breached pairing allows all
flavors to form condensates $\langle ud \rangle_{BP}, \langle us
\rangle_{BP}, \langle ds \rangle_{BP}$ of breached pairing type.  We
suspect that there is a density range where it is energetically
favorable compared to unpaired matter or the $2$SC phase with only
$\langle ud \rangle_{BCS} \neq 0$ condensation.  Alford and Rajagopal
\cite{Alford:02} showed the absence of $2$SC phase in compact stars,
and suggested that there should be some non-BCS pattern of pairing
including possible unpaired matter in the range of densities between
hadronic and CFL matter.  The breached pairing version of CFL appears
to be a serious candidate for these intermediate densities.  Since
there is a simple second order phase transition from BP to BCS
superfluidity, as a function of increasing density, it is natural to
expect a single phase transition from breached pairing CFL to standard CFL
quark matter.  Detailed calculations, including shifts of Fermi
surfaces due to electric and color neutrality, will be required to
determine the range, if any, over which breached pairing CFL is favored
(in preparation).  
At strong coupling one could even speculate about breached pairing CFL
states at zero strangeness, where the strange chemical potential falls
between $\pm m_s$.

The possibility that one might rearrange the free-particle Fermi
surfaces to take advantage of opportunities for pairing was considered
long ago in condensed matter literature by Sarma and others
\cite{Sarma:63+TI:69}. 
Superfluidity with quasiparticle dispersion relation
crossing zero, thus leading to gapless modes, was encountered in a
model resembling ours in Ref.~\cite{ABR:00}. In neither case, however, was
the possibility of a {\it stable\/} breached pairing state recognized. (See
also a recent study in cold atoms \cite{Wu-Yip:03}.)  
As this work was being completed reference Ref.~\cite{Shovkovy:03} appeared, 
in which a related problem, imposing a charge neutrality condition directly, 
was analyzed.

We have phrased our discussion in terms of QCD and quarks, but one can
easily imagine closely related universality classes for condensed
matter or cold atom systems.

The authors thank Jeff Bowers for useful discussions and valuable
suggestions, Wolfram Schroers for help in numerical analyses; and
M. Forbes, J. Kundu, Y. Schroeder and I. Sigalov. This work is
supported in part by funds provided by the U.S.Department of Energy
(D.O.E.) under cooperative research agreement DF-FC02-94ER40818.

\bibliographystyle{apsrev}
\bibliography{breach_pair}

\end{document}